# Collector Failures on 350 MHz, 1.2 MW CW Klystrons at the Low Energy Demonstration Accelerator (LEDA)

D. Rees, W. Roybal, J. Bradley III, LANL, Los Alamos, NM 87545, USA

*Abstract*

We are currently operating the front end of the accelerator production of tritium (APT) accelerator, a 7 MeV radio frequency quadrupole (RFQ) using three, 1.2 MW CW klystrons. These klystrons are required and designed to dissipate the full beam power in the collector. The klystrons have less than 1500 operational hours. One collector has failed and all collectors are damaged. This paper will discuss the damage and the difficulties in diagnosing the cause. The collector did not critically fail. Tube operation was still possible and the klystron operated up to 70% of full beam power with excellent vacuum. The indication that finally led us to the collector failure was variable emission. This information will be discussed. A hydrophonic system was implemented to diagnose collector heating. The collectors are designed to allow for mixed-phase cooling and with the hydrophonic test equipment we are able to observe: normal, single-phase cooling, mixed-phase cooling, and a hard boil. These data will be presented. The worst case beam profile from a collector heating standpoint is presented. The paper will also discuss the steps taken to halt the collector damage on the remaining 350 MHz klystrons and design changes that are being implemented to correct the problem.

## 1 INTRODUCTION

The APT 350 MHz, 1.2 MW, CW klystrons operating on LEDA are similar in design to the English Electric Valve (EEV) 352 MHz klystrons at CERN, APS, and ESRF with one difference. The APT klystrons were designed to allow for the steady-state dissipation of the full DC beam power (95 kV, 20 A, 1.9 MW) in the collector. This requirement was intended to mitigate AC grid transients which could result from loss of accelerator beam since the APT accelerator is heavily beam loaded or from interlock-induced, short-term, accelerator outages. Each klystron was factory tested to this requirement and operated at LEDA without an interlock to limit the time the collector was subject to the full beam power in the event the RF drive to the klystron is interrupted.

Three klystrons of this type provide power to a RFQ. Approximately 2.4 MW of RF power is required from the klystrons for the RFQ. Three klystrons are connected to the RFQ which acts as the power combiner. The system is designed so that only two of the three klystrons are required for operation. Waveguide switches are included in each waveguide run so that a failed klystron can be removed. The switch reflects a short circuit at the appropriate phase back to the RFQ and the other two klystrons can be used to continue operations. A picture of the klystron is included in Fig. 1.

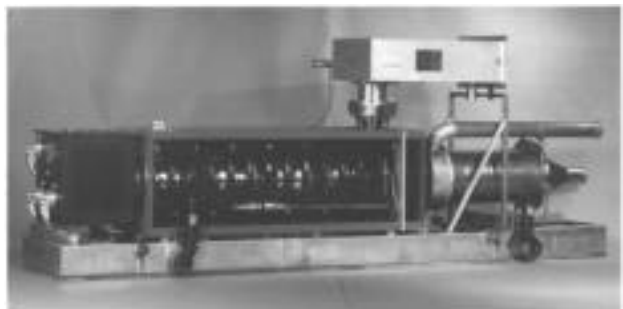

Figure 1: EEV 350 MHz APT klystron.

## 2 DISCOVERY OF DAMAGE

### 2.1 Initial Indications

The first indication of a problem was an audible boiling in the collector. The collector is a mixed-phase collector which allows for boiling and recondensation within the collector grooves. The audible boiling was not this mixed-phase boiling. The noise was the result of large steam bubbles being transported at the high water velocities through the collector cooling pipes and metering. We notified the vendor of the boiling and were told it was nothing to worry about. This boiling was noticeably worse on one of the three klystrons, but was observable on all klystrons for high collector powers.

In order to keep the audible boil to a minimum we had been operating the klystrons at the minimum beam power required for RFQ operations. Earlier in the day we had increased the klystron beam power from approximately 1.1 MW to 1.5 MW to allow for increases in RFQ beam current and provide more margin for field control. We had operated several hours at this new operating point with negligible ion pump current (less than .2 uA) when we experience a large vacuum event that triggered both the high voltage interlock (which triggers at 10 uA ion pump current) and the filament power interlock (which triggers at 100 uA of ion pump current). We suspended operations and attempted to recondition the klystron. Although we were able to recondition the klystron, as we operated at higher beam power we began to observe variable cathode emission.

### 2.2 Variable Emission

As we reconditioned the klystron, we discovered that as we increased beam power, cathode current would vary by several amps when increasing or decreasing the RF drive. We could also cause this same variation by small changes in cathode voltaage. Our system operates with a mod-anode regulator tube that maintains the mod-anode voltage at a fixed setpoint relative to the cathode voltage and we carefully monitored this voltage, as well as the filament current, to verify neither of these factors were causing the change in cathode current. Finally, we did a series of tests where we triggered our arc detectors which inhibit RF drive to the klystrons.

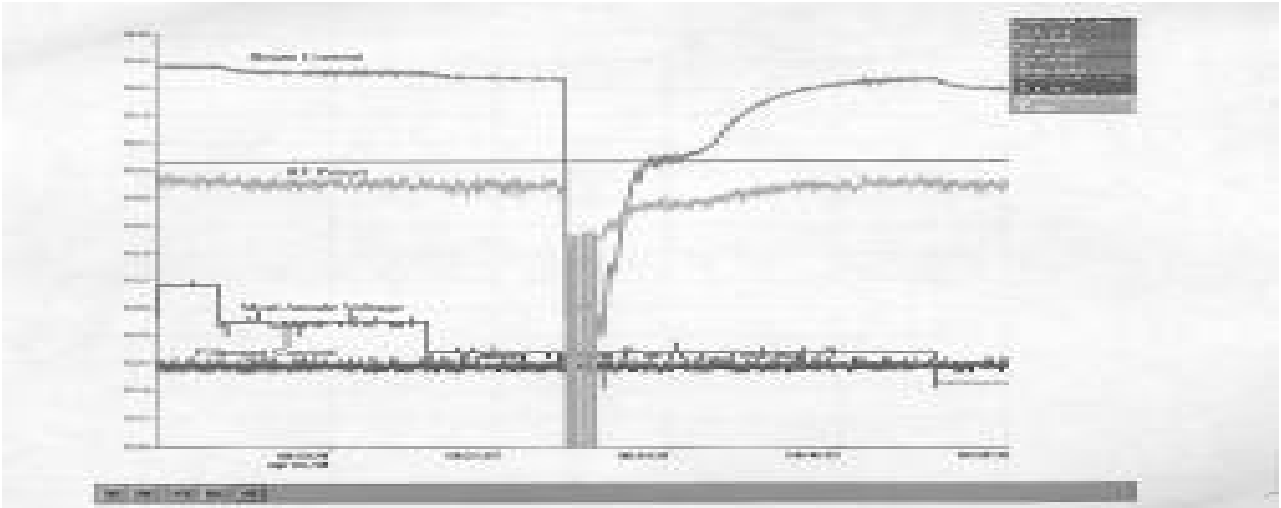

Figure 2: Strip chart record illustrating variable emission with changes in collector power.

Fig. 2 shows the results of one of these tests. In the figure, the RF power is observed to go to zero and a coresponding plunge in cathode current is observed. When the RF power is reestablished, the cathode current is observed to slowly return to its previous value. The filament current and mod-anode voltage relative to cathode voltage is observed to be constant over the variation in cathode current. From this data we concluded copper vapor from the collector was poisoning the cathode emission as we increased the thermal load in the collector. It was also interesting that the vac-ion pump power supply did not register an increase in vacuum and indicated very low current throughout this test.

## 2.3 Damage Assessment

Based on this result and conversations over the duration of our testing, EEV had two klystron engineers visit our site. We removed the end of the collector water jacket from the klystron and observed the damage shown in Fig. 3. From Fig. 3 it is observed that several portions of the collector had become quite hot and dimples had resulted from a combination of the heat, water pressure, and vacuum forces. The general collector shape had also distorted. It was now elliptical in cross section rather than round. Also, the collector end had drooped and the bottom of the conical shaped portion of the collector was actually resting on the water jacket.

## 2.4 Short Term Solution

We inspected the collectors of all the klystrons and all showed various levels of damage ranging from minimal to the significant damage shown in Fig. 3. All collectors showed indications of droop and three of the four collectors were becoming elliptical rather than round in cross section. All collectors had less than 1000 high voltage hours and one had only 60 high voltage hours since delivery.

Of the four klystrons, one klystron had a bucking coil in place to minimize magnetic field penetration into the collector. This klystron had the least damage. Two other klystrons had this bucking coil installed but not electrically connected to the magnetic circuit. The fourth klystron did not have the bucking coil installed. We were dismayed to discover the variation in the magnetic circuit of the klystrons, but were encouraged by the fact that the klystron with the bucking coil operable showed the least damage. We took the following short-term steps to return the klystrons to service due to programmatic pressure.

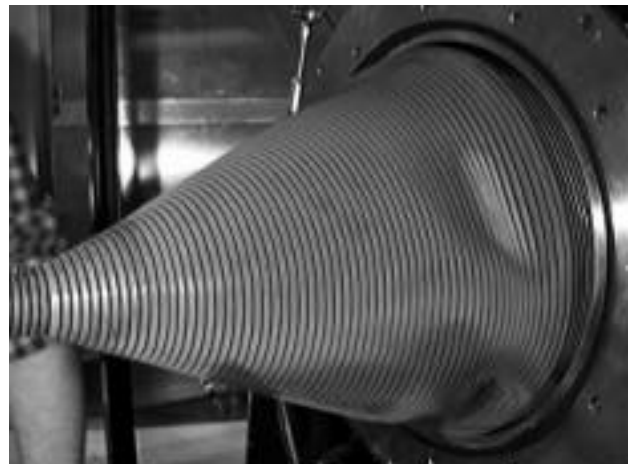

Figure 3: External view of collector damage.

The klystron in Fig. 3 was returned to the factory for rebuild still under vacuum and operable at low beam powers. The collector jackets of the remaining three klystrons were adjusted to center the collectors in the water jackets in order to compensate for the droop, and supports were welded on the inside of the water jacket to support the end of the collector to avoid further droop. An acoustical measurement system was installed on the collector to validate the effect of the bucking coil and to characterize safe regions of operation. The sensor was a microphone on the collector cooling return. These results are presented in Fig. 4. The four curves in Fig. 4 show some unexpected results. The mixed phase boiling that the collector was designed to accommodate manifests as an increase in the higher frequency audio components of Fig. 4. This increase can be used as a qualitative indication of peak collector heat load. From Fig. 4 it can be seen that the localized collector heat load at 77 kV, 14.5 A is actually higher than at 85 kV, 17.1 A. This suggests that the beam power is better distributed over the collector at 85 kV. The plots in Fig. 4 were measured without RF power with the full beam power dissipated in the collector. Fig. 4 also shows that application of the bucking coil lowers the localized collector heat load. Using this tool we were able to define an allowable safe operating region and also regions to avoid. Operation with three klystrons at an 85 kV voltage provides sufficient power for the RFQ with full accelerator beam current and sufficient control margin.

The data in Fig. 4 were measured on the klystron that had the second most significant amount of collector damage. The other two klystrons showed a similar behavior but not the same magnitude of accoustical variation. For all klystrons, operation between 70 kV and 80 kV resulted in audio indications of increased localized heating.

In addition to the steps taken above, we also implemented a beam interlock such that above 70 kV of beam voltage, 100 kW of RF power was required or an interlock would remove high voltage within 15 seconds.

We would have preferred a shorter interlock period, but the 15 second interval was a compromise with the RFQ operations team and with EEV.

## 3 LONG TERM SOLUTIONS

EEV has redesigned the collector so it is capable of achieving the initial requirements. The details provided below are intentionally vague to protect proprietary information. Additional modeling was conducted using a recently released, new version of one of the beam dynamics codes originally used in the collector design. The beam was discovered to penetrate further into the collector than originally thought. Additional length has been added to the collector in the new collector design some. The conical collector end has been changed to a shape which reduces the thermal stress. The bucking coil is now a required part of the design. The water jacket design has been changed to maintain a high flow velocity over the entire collector. The supports on the end of the collector implemented as part of the short term solution have become a part of the design. We also conducted x-ray measurements of the klystron internal to the lead shielding and discovered that the beam alignment could be improved. These results were duplicated by EEV. EEV is taking steps to improve beam alignment.

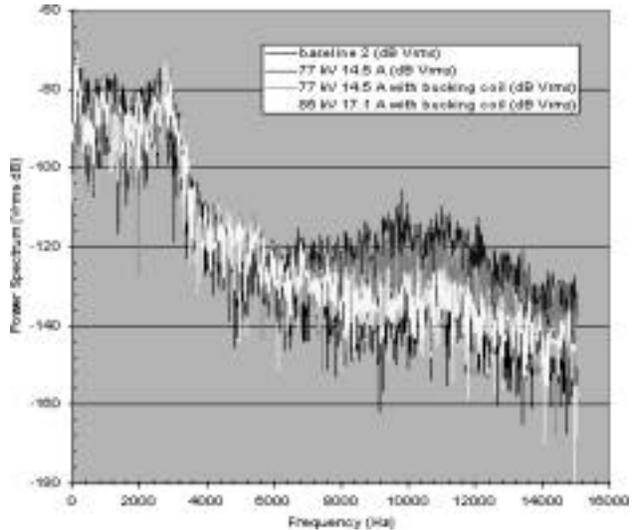

Figure 4: Collector audio profile.

## 4 CONCLUSION

We discovered the initial collector design to be inadequate for satisfying our requirements. We have worked closely with EEV to diagnose and modify the collector design. The first 350 MHz klystron with the new collector design is scheduled for factory testing in August, 2000. The factory acceptance tests of the collector have been significantly increased and an inspection of the collector is conducted at the conclusion of the testing. After delivery of this initial tube, another klystron from Los Alamos will be rebuilt to include the new collector. The remaining two klystrons at Los Alamos had only slight collector damage and will continue to be used on LEDA.

As part of diagnosing the collector problem, we developed a useful and interesting tool to characterize localized collector heat load using an acoustical measurement. This tool led to the unexpected result that lower beam powers had higher localized collector heat loads than operation a higher beam powers.

We have operated the three remaining klystrons for slightly over 1000 cumulative hours since the collector problem was discovered. The two slightly damaged klystrons show no sign of further damage. The remaining klystron with the most significant damage is showing signs of further degradation. We hope that it will continue to meet the RFQ needs until the replacement tube arrives in early September.